\documentclass[10pt,onecolumn] {IEEEtran}

\usepackage{graphicx}
\usepackage{float}
\usepackage{amsthm}
\usepackage{amsmath}
\usepackage{amsfonts}
\usepackage{listings}
\usepackage{paralist}
\usepackage{cite}

\theoremstyle{plain}\newtheorem{thm}{Theorem}
\theoremstyle{definition}\newtheorem{defn}{Definition}
\theoremstyle{plain} \newtheorem{cor}{Corollary}
\theoremstyle{remark} \newtheorem{ex}{Example}
\newcommand{\Frac}[2]{{{#1}/{#2}}}  
\DeclareMathOperator{\E}{E}
\newcommand{\Quant}{Q_{K,\lambda}}  
\newcommand{\QuantV}{Q_{K_1^N,\lambda_1^N}}
  
\newcommand{\Lv}{\lambda_1^N}  
\newcommand{\Kv}{K_1^N}  
\newcommand{\D}{D_\mathrm{fmse}(K,\lambda)}
\newcommand{\Dv}{D_\mathrm{fmse}(K_1^N,\lambda_1^N)}
\def\APPROX{\simeq}

\newcommand{\beq}{\begin{equation}}
\newcommand{\eeq}{\end{equation}}
\newcommand{\beqa}{\begin{eqnarray}}
\newcommand{\eeqa}{\end{eqnarray}}
\newcommand{\beqan}{\begin{eqnarray*}}
\newcommand{\eeqan}{\end{eqnarray*}}

\usepackage{color}
\usepackage[normalem]{ulem} 

\begin{document}

\title{Distributed Functional Scalar Quantization Simplified}

\author{John~Z.~Sun$^*$,~\IEEEmembership{Student~Member,~IEEE,} 
Vinith~Misra,~\IEEEmembership{Student~Member,~IEEE,}
and~Vivek~K~Goyal,~\IEEEmembership{Senior~Member,~IEEE}%
\thanks{
This material is based upon work supported by the National Science Foundation
under Grants 0643836, 0729069, and 1115159\@.}%
\thanks{J. Z. Sun and V. K. Goyal are with the Department of Electrical Engineering and Computer
Science and the Research Laboratory of Electronics, Massachusetts Institute of
Technology, Cambridge, MA 02139 USA (e-mail: \{johnsun,vgoyal\}@mit.edu).}
\thanks{V. Misra is with the Department of Electrical Engineering and the Information Systems Laboratory, Stanford University, Stanford, CA 
94305 USA (e-mail: vinith@stanford.edu).}}

\maketitle
\begin{abstract}
		Distributed functional scalar quantization (DFSQ) theory provides optimality conditions and predicts 
		performance of data acquisition systems in which a computation on acquired data is desired.
		We address two limitations of previous works: prohibitively expensive decoder design and a restriction to sources with
		bounded distributions. 
		We rigorously show that a much simpler decoder has equivalent
		asymptotic performance as the conditional expectation estimator previously explored,
		thus reducing decoder design complexity.  The simpler decoder has the feature of decoupled
		communication and computation blocks.
		Moreover, we extend the DFSQ framework with the simpler decoder to acquire sources with
		infinite-support distributions such as Gaussian or exponential distributions. 
		Finally, through simulation results we demonstrate that performance at moderate coding rates is well 
		predicted by the asymptotic analysis, and we give new insight on the rate of convergence.
\end{abstract}

\section{Introduction}
\label{sec:intro}

\begin{figure}
  \begin{center}
    \includegraphics[width=3in]{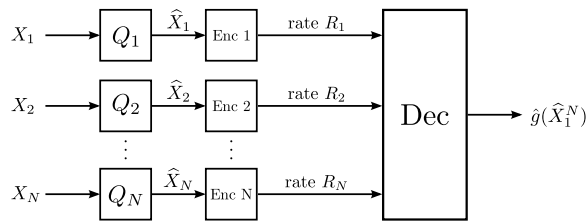}
  \end{center}
  \caption{A distributed computation network, where each of $N$ spatially-separated sources generate a scalar 
  		$X_n$. 
			The scalars are encoded and communicated over rate-limited links to a central decoder without 
			interaction between encoders.
			The decoder computes an estimate of the function $g(X_1^n) = g(X_1,\,X_2,\,\ldots,\,X_n)$ 
			from the received data using $\hat{g}(\widehat{X}_1^n)$.
			Each encoder is allowed transmission rate $R_n$. }
  \label{fig:star}
\end{figure}

\PARstart{F}{unctional} source coding techniques are of great importance in modern
distributed systems such as sensor networks and cloud computing architectures
because the fidelity of acquired data can greatly impact the accuracy
of computations made with that data.
In this work, we provide theoretical and empirical results for quantization in 
distributed systems described by the topology in Fig.~\ref{fig:star}. 
Here, $N$ memoryless sources produce scalar realizations $X_1^N = (X_1 , \ldots , X_N)$ 
from a joint distribution $f_{X_1^N}$ at each discrete time instant. 
These measurements are compressed via separate encoders and then sent to a 
central decoder that approximates a computation on the original data;
the computation may be the identity function, meaning that the acquired
samples themselves are to be reproduced. 

There has been substantial effort to study distributed coding using information-theoretic concepts, 
taking advantage of large block lengths and powerful decoders to approach fundamental limits of compression. 
However, techniques inspired by this theory are infeasible for most applications.  In particular, strong dependencies between source variables imply low
information content per variable, but exploiting this is difficult under
rigid latency requirements.

Rather than have long blocks,
the complementary asymptotic of
\emph{high-resolution quantization theory}~\cite{Neuhoff:93} is more 
useful for these scenarios; 
most of this theory is focused on the scalar case, where the block length is one.
The principal previous work in applying high-resolution quantization theory
to the acquisition and computation network of Fig.~\ref{fig:star} is the
\emph{distributed functional scalar quantization} (DFSQ)
framework~\cite{MisraGV:11}.
The key message from DFSQ is that the design of optimal encoders for systems
that perform nonlinear computations can be drastically different from 
what traditional quantization theory suggests.
In recent years, ideas from DFSQ have been applied to
compressed sensing~\cite{SunG:09-ISIT},
compression for media~\cite{SunG:11},
and channel state feedback in wireless networks~\cite{PughR:11}. 

Like the information-theoretic approaches, the existing DFSQ theory
relies in principle on a complicated decoder.
(This reviewed in Section~\ref{sec:prelim:hrfsq}.)
The primary contribution of this paper is to study a
DFSQ framework that employs a simpler decoder.
Remarkably,
the same asymptotic performance is obtained with the simpler decoder,
so the optimization of quantizer point density is unchanged.
Furthermore, the simplified framework allows a greater decoupling or
modularity between communication (source encoding/decoding) and
computation aspects of the network.

The analysis presented here uses different assumptions on the source
distributions and function than~\cite{MisraGV:11}---neither is uniformly
more or less restrictive.  Unlike in~\cite{MisraGV:11},
we are able to allow the source variables to have infinite support.
In fact, the functional setting allows us to
present high-resolution quantization results for certain heavy-tailed
source distributions for the first time.

We begin in Sec.~\ref{sec:prelim} by reviewing relevant previous work and summarizing the contributions
of this paper. 
In Sec.~\ref{sec:uv} and~\ref{sec:mv}, we give distortion and design results for a distributed network.
Finally, we provide examples for the theory in Sec.~\ref{sec:examples} and 
conclude in Sec.~\ref{sec:summary}.

\section{Preliminaries}
\label{sec:prelim}

\subsection{Previous Work}
\label{sec:prelim:prev}

The distributed network shown in Fig.~\ref{fig:star} is of great interest to the information theory and 
communications communities, and there exists a variety of results corresponding to different scenarios of interest. 
We present a short overview of some major works; a comprehensive review appears in~\cite{MisraGV:11}.

In the large block length asymptotic, there are many influential and conclusive results. 
For the case of discrete-valued sources
and $g(X_1^N) = X_1^N$, the lossless distributed source coding problem is 
solved by Slepian and Wolf~\cite{SlepianW:73}.
In the lossy case, the problem is generally open except in specific situations~\cite{WagnerTV:08,ZamirB:99}.
The case where $g(X_1^N) = X_1$ and the rate is unconstrained except for $R_1$ is the well-known
source coding with side information problem~\cite{WynerZ:76}. 
For more general computations, the lossless~\cite{OrlitskyR:01,HanK:87,DoshiSMJ:07} and
lossy~\cite{Yamamoto:82,FengES:04} cases have both been explored. 

There are also results for when the block length is constrained to be very small. 
We will delay discussion of DFSQ for later and instead focus on related works. 
The use of high-resolution for computation has been considered in detection and estimation 
problems~\cite{Poor:88,BenitzB:89,GuptaH:03}. 
In the scalar setting, the scenario where the computation is unknown but is drawn from a set
of possibilities has been studied~\cite{Bucklew:84}. 
Finally, there are strong connections between DFSQ and multidimensional companding, a technique
used in perceptual coding~\cite{LinderZZ:99}.

\subsection{High-resolution Scalar Quantizer Design}
\label{sec:prelim:hrsq}

A scalar quantizer $Q_K$ is a mapping from the real line to a set of $K$ points
$\mathcal{C} = \{c_k\}_{k=1}^K \subset \mathbb{R}$ called the codebook,
where $Q_K(x) = c_k$ if $x \in P_k$ and the cells
$\{ P_k \}_{k=1}^K$ form a partition of $\mathbb{R}$. 
The quantizer is called \emph{regular} if the partition cells are intervals containing the corresponding codewords.
We then assume the codebook entries are indexed from smallest to largest
and that $P_k = (p_{k-1}, p_k]$ for each $k$;
this is essentially without loss of generality because the dispositions
of the endpoints of the cells are immaterial to performance when the
quantizer input is continuous.
Regularity implies $p_0 < c_1 \leq p_1 < c_2 \leq \cdots < c_K \leq p_K$,
with $p_0 = -\infty$ and $p_K = \infty$. 
Define the \emph{granular} region as $(c_1,c_K)$ and its complement
$(-\infty,c_1] \cup [c_K,\infty)$ as the \emph{overload} region. 

\begin{figure}
  \begin{center}
    \includegraphics[width=3.2in]{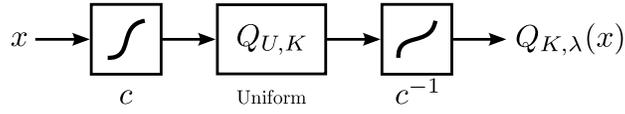}
  \end{center}
  \caption{A block diagram for companding as a constructive method for non-uniform scalar quantization.
  		The notation $Q_{U,K}$ is used to describe the canonical uniform quantizer with $K$ partitions in the 
  		granular region $[0,1]$.
  		In this paper, only the partition boundaries are scaled using compressor function $c$; 
  		the codewords are defined through
  		midpoint reconstruction~\eqref{eq:midpointReconstruction} and can be computed at the decoder.}
  \label{fig:compander}
\end{figure}

Uniform (linear) quantization, where partition cells in the granular region have equal length,
is most commonly used in practice, but other quantizer designs are possible. 
Fig.~\ref{fig:compander} presents the compander model as a method for generating 
nonuniform quantizers from a uniform one.
In this model, the scalar source is transformed using a nondecreasing and 
smooth
\emph{compressor} function $c : \mathbb{R} \rightarrow [0,1]$, 
then quantized using a uniform quantizer comprising $K$ levels on the granular region $[0,1]$, 
and finally passed through the \emph{expander} function $c^{-1}$.
Compressor functions are defined such that $\lim_{x\rightarrow-\infty} c(x) = 0$ and
$\lim_{x\rightarrow\infty} c(x) = 1$.
It is convenient to define a \emph{point density function} as $\lambda(x) = c^\prime(x)$.
Because of the extremal conditions on $c$, there is a one-to-one correspondence between $\lambda$ and $c$,
and hence a quantizer of the form shown in Fig.~\ref{fig:compander} can be uniquely specified using a point 
density function and codebook size. 
We denote such a quantizer as $\Quant$.
By virtue of this definition, the integral of the point density function over
any quantizer interval is $1/K$:
\beq
\label{eq:densityKRelation}
\int_{p_k}^{p_{k+1}} \lambda(x) \, dx = \frac{1}{K} \mbox{,}
\qquad k = 1,\,2,\,\ldots,\,K \mbox{.}
\eeq

In practice, scalar quantization is rarely, if ever, performed by an explicit companding operation.  
A slight modification that avoids repeated computation of $c^{-1}$
is to apply the compressor $c$, compare to threshold values
(multiples of $1/K$) to determine the partition cell $P_k$,
and then obtain $c_k$ from a pre-computed table.
We assume that the non-extremal reconstruction values are set to the midpoints of the cells, i.e.\
\beq
\label{eq:midpointReconstruction}
c_k = \frac{p_{k-1} + p_k}{2} \mbox{,}
\qquad k = 2,\,3,\,\ldots,\,K-1 \mbox{.}
\eeq
This is suboptimal relative to centroid reconstruction,
but has the simplicity of depending only on $\lambda$ and $K$---not on
the source density.
The extremal reconstruction values are fixed to be $c_1 = p_1$ and $c_K = p_{K-1}$.
This again is suboptimal but does not depend on the source distribution.
We will show later that this construction does not affect asymptotic quantizer performance.

The utility of the compander model is that we can precisely analyze the
distortion behavior as $K$ becomes large and use this to optimize $\lambda$.
Assuming the source is well-modeled as being drawn iid from a probabilistic distribution, 
we define the mean-squared error (MSE) distortion as
\beq
	\label{eq:mse:eq}
	D_\mathrm{mse}(K,\lambda) = \E [ |X - \Quant(X)|^2 ] ,
\eeq
where the expectation is with respect to the source density $f_X$.
Under the additional assumption that $f_X$ is continuous (or simply measurable) 
with tails that decay sufficiently fast,
\beq
	\label{eq:mse:approx}
	D_\mathrm{mse}(K,\lambda) \APPROX \frac{1}{12 K^2} \E [ \lambda^{-2}(X) ] ,
\eeq
where $\APPROX$ indicates that the ratio of the two expressions approaches 1 
as $K$ increases~\cite{Bennett:48,PanterD:51}.
Hence, the MSE performance of a scalar quantizer can be approximated by a simple
relationship between the source distribution, point density and codebook size, and this
relation becomes more precise with increasing $K$. 
Moreover, quantizers designed according to this approximation are \emph{asymptotically optimal}, meaning that
the quantizer optimized over $\lambda$ has distortion that approaches the performance of the best $Q_K$
found by any means~\cite{BucklewW:82,CambanisG:83,Linder:91}, meaning
\beq
	\label{eq:mse:asymptoticallyoptimal}
	\inf_{Q_K} \E \left[ |X -  Q_K(X) |^2 \right] \APPROX \frac{1}{12 K^2} \E [ \lambda^{-2}(X) ] .
\eeq
Experimentally, the approximation is accurate even for moderate $K$~\cite{Neuhoff:93,Goyal:00b}.
Since distortion depends only on $\lambda$ in the asymptote, calculus techniques
can be used to optimize companders. 

When the quantized values are to be communicated or stored, it is natural to map codewords to 
a string of bits and consider the trade-off between performance and communication rate $R$, 
defined to be the expected number of bits per sample.
In the simplest case, the codewords are indexed and the communication rate is $R = \log_2(K)$; 
this is called \emph{fixed-rate} or \emph{codebook-constrained} quantization. 
H\"{o}lder's inequality can be used to show that the optimal point density for fixed-rate is 
\begin{equation}
	\label{eq:mse:lamfr}
	\lambda^*_{\mathrm{mse,fr}}(x) \propto f_X^{1/3}(x) ,
\end{equation}
and the resulting distortion is 
\begin{equation} 
	\label{eq:mse:fr:distopt}
	D^*_\mathrm{mse,fr}(R) \APPROX \frac{1}{12} \, \| f_X \|_{1/3} \, 2^{-2R},
\end{equation}
with the notation $\|f\|_p = (\int_{-\infty}^\infty f^p(x) \, dx)^{1/p}$~\cite{GrayG:77}.

In general, the codeword indices can be coded to produce bit strings of different lengths based
on probabilities of occurrence;
this is referred to as \emph{variable-rate} quantization.
If the decoding latency is allowed to be large, one can employ block entropy coding and the 
communication rate approaches $H(\Quant(X))$.
This particular scenario, called \emph{entropy-constrained} quantization, can be analyzed using Jensen's
inequality to show the optimal point density $\lambda^*_\mathrm{mse,ec}$ is constant 
on the support of the input distribution~\cite{GrayG:77}.
The optimal quantizer is uniform and the resulting distortion is
\begin{equation} 
	\label{eq:mse:vr:distopt}
	D^*_\mathrm{mse,ec}(R) \APPROX \frac{1}{12} 2^{-2(R-h(X))} .
\end{equation}
Note that block entropy coding suggests that the sources are transmitted in blocks even though 
the quantization is scalar.
As such, \eqref{eq:mse:vr:distopt} is an asymptotic result and serves as a lower bound on practical entropy 
coders with finite block lengths that match the latency restrictions of a system.

In general, the optimal entropy-constrained quantizer (at a finite rate) for a distribution with unbounded support
can have an infinite number of codewords~\cite{GyorgyLCB:03}.
The compander model used in this paper cannot generate all such quantizers.
A common alternative is to allow the codomain of $c$ to be $\mathbb{R}$
rather than $[0,1]$, 
resulting in a point density that cannot be normalized~\cite{GishP:68,GrayN:98}.
To avoid parallel developments for normalized and unnormalized point densities,
we restrict our attention to quantizers that have a finite number of codewords
$K$ at any finite rate $R$.  This may preclude exact optimality,
but it does not change the asymptotic behavior as $K$ and $R$ increase
without bound.
Specifically, the contribution to overall distortion from the
overload region is made negligible as $K$ and $R$ increase,
so the distinction between having finitely- or infinitely-many codewords
becomes unimportant.

\subsection{Functional Scalar Quantizer Design}
\label{sec:prelim:hrfsq}

In a distributed network where the encoders employ scalar quantization and the decoder performs
a known computation, 
optimizing for the computation rather than source fidelity can lead to substantial gains. 
In~\cite{MisraGV:11}, distortion performance and quantizer design are discussed for the distributed setting
shown in Fig.~\ref{fig:star}, with $g$ a scalar-valued function.
For DFSQ, the cost of interest is functional MSE (fMSE):
\begin{equation}
	\label{eq:fmse} 
	D_\mathrm{fmse}(\Kv, \Lv) = \E \left[ |g(X_1^N) - \hat{g}(\QuantV(X_1^N)) |^2 \right] ,
\end{equation}
where $g$ is a scalar function of interest, $\hat{g}$ is the optimal fMMSE estimator
\begin{equation}
	\label{eq:optdec}
	\hat{g}(x_1^N) = \E \left[ g(X_1^N) \, | \, \QuantV(X_1^N) = \QuantV(x_1^N) \right] ,
\end{equation}
and $\QuantV$ is scalar quantization performed on a vector such that
\[
\QuantV(x_1^N) = \left( Q_{\lambda_1, K_1}(x_1) , \ldots Q_{\lambda_N, K_N}(x_N) \right) \mbox{.}
\]
Note the complexity of computing $\hat{g}$:
it requires integrating over an $N$-dimensional partition cell
with knowledge of the joint source density $f_{X_1^N}$.
Later in this paper, we avoid this complexity by setting $\hat{g}$ to equal $g$.

Before understanding how a quantizer affects fMSE, it is convenient to define how a computation locally 
affects distortion. 
\begin{defn}
	\label{def:sens}
	The {\it univariate functional sensitivity profile} of a function $g$ is defined as 
	\[
	 \gamma(x) = |g^\prime(x)|.
	\]
	The {\it $n$th functional sensitivity profile} of a multivariate function $g$ is defined as 
	\beq
		\label{eq:gamman}
		\gamma_n(x) = \left( \E \left[ |g_n(X_1^N)|^2 \, | \, X_n = x \right]\right)^{1/2} ,
	\eeq
	where $g_n(x)$ is the partial derivative of $g$ with respect to its $n$th argument evaluated at the point $x$.
\end{defn}

Given the sensitivity profile, the main result of~\cite{MisraGV:11} says
\begin{equation}
	\label{eq:fmse:dist} 
	D_\mathrm{fmse}(\Kv, \Lv) \APPROX \sum_{n=1}^N \frac{1}{12 K_n^2} 
					\E \left[ \left( \frac{\gamma_n(X_n)}{\lambda_n(X_n)} \right)^2 \right] ,
\end{equation}
provided the following conditions are satisfied:
\begin{asparaenum}[MF1.]
	\item The function $g$ is Lipschitz continuous and twice differentiable in every argument
					except possibly on a set of Jordan measure 0. 
	\item The source pdf $f_{X_1^N}$ is continuous, bounded, and supported on $[0,1]^N$.
	\item The function $g$ and point densities $\lambda_n$ allow
					$\E[( \Frac{\gamma_n(X_n)}{\lambda_n(X_n)} )^2 ]$ to be defined and finite for all $n$. 
\end{asparaenum}

Following the same recipes to optimize over $\Lv$, the relationship between
distortion and communication rate is found.
In both cases, the sensitivity acts to shift quantization points to where they can reduce the 
distortion in the computation. 
For fixed rate,  the minimum high-resolution distortion is achieved by
\begin{equation}
	\label{eq:dfsq:lamfr}
	\lambda^*_{n,\mathrm{fmse,fr}}(x) \propto \left( \gamma_n(x) f_{X_n}(x) \right)^{1/3} ,
\end{equation}
where $f_{X_n}$ is the marginal distribution of $X_n$. 
In the entropy-constrained case, the optimizing point density is 
\begin{equation}
	\label{eq:dfsq:lamvr}
	\lambda^*_{n,\mathrm{fmse,ec}}(x) \propto \gamma_n(x) .
\end{equation}
Notice unnormalized point densities are not required here since the source is assumed to have bounded support.

\subsection{Main Contributions of Paper}
\label{sec:prelim:contribution}

The central goal of this paper is to develop a more practical method upon the theoretical foundations of~\cite{MisraGV:11}.
In particular,
we provide new insight on how a simplified decoder can be used in lieu of the optimal one in~\eqref{eq:optdec}.
Although the conditional expectations are offline computations, they may be extremely difficult
and are computationally infeasible for large $N$ and $K$.
We consider the case when the decoder is restricted to
applying the function $g$ explicitly on the quantized measurements.  
To accommodate this change and provide more intuitive proofs,
a slightly different set of conditions is required of $g$, $\lambda_1^N$, and $f_{X_1^N}$.

Additionally, we generalize the theory to infinite-support source variables
and vector-valued computations. 
In brief, we derive new conditions on the tail of the source density and computation that allow
the distortion to be stably computed. 
Interestingly, this extends the class of probability densities under
which high-resolution analysis techniques have been successfully applied.
The generalization to vector-valued $g$ is a more straightforward extension
that is included for completeness.
We present several examples to illustrate the framework and the convergence
to the asymptotics developed in this work.

\section{Univariate Functional Quantization}
\label{sec:uv}

We first discuss the quantization of a scalar random variable $X$ by $\Quant$,
with the result fed into a function $g$ in order to approximate $g(X)$. 
As mentioned, this is a simpler decoder than analyzed in~\cite{MisraGV:11}.
We find the dependence of fMSE on $\lambda$ and then optimize with respect to $\lambda$ to minimize fMSE\@. 

Assume a companding quantizer with point density $\lambda$ and granular region $S_K \subset \mathbb{R}$
Consider the following conditions on the computation $g$ and the density $f_X$ of the source: 
\begin{asparaenum}[UF1$^\prime$.]
	\item The source pdf $f_X$ is continuous and strictly positive on $S_K$ for any finite $K$. 
	\item The function $g$ is continuous on $S_K$ with both $|g^\prime|$ and $|g^{\prime\prime}|$ 
					defined and bounded by a finite constant $C_u$.
	\item $f_X(x) |g^\prime(x)|^{2-m} / \lambda^{2+m}(x)$ is Riemann integrable over $S_K$ for $m=0,1,2$.
	\item $f_X$, $g$ and $\lambda$ satisfy the tail condition 
			\[ \lim_{y \to \infty} \frac{ \int_y^\infty |g(x) - g(y)|^2 f_X(x) \, dx}
						{\left( \int_y^\infty \lambda(x) \, dx \right)^2} = 0 , \]
				and the corresponding condition for $y \to -\infty$.
\end{asparaenum}

The main result of this section is on the fMSE induced by a quantizer $\Quant$ under these conditions:
\begin{thm}
	\label{thm:uvdist}
	Assume $f_X$, $g$, and $\lambda$ satisfy conditions UF1$^\prime$--UF4$^\prime$. Then the fMSE
	\begin{equation}
		\label{eq:uvdist}
		\D = \E \left[ | g(X) - g(\Quant(X)) |^2 \right]
	\end{equation}
	has the following limit:
	\begin{equation}
		\label{eq:uvdistHR}
		\lim_{K \to \infty} K^2 \D 
				= \frac{1}{12} \E \left[ \left(\frac{\gamma(X)}{\lambda(X)} \right)^2 \right] .
	\end{equation}
\end{thm}

\begin{IEEEproof}
	See Appendix~\ref{app:thm:uvdist}.
\end{IEEEproof}

\subsection{Remarks}
\label{sec:uv:remarks}

\begin{asparaenum}[1.]
	\item The fMSE in \eqref{eq:uvdistHR} is the same as in \eqref{eq:fmse:dist}.
			We emphasize that the theorem shows that this fMSE is obtained by simply
			applying $g$ to the quantized variables rather than using the optimal decoder
			$\hat{g}$ from \eqref{eq:optdec}. 
			Further analysis on this point is given in Sec.~\ref{sec:uv:subopt}.
	\item When $g$ is monotonic, the performance \eqref{eq:uvdistHR} is as good as
			quantizing and communicating $g(X)$~\cite[Lemma 5]{MisraGV:11}.
			Otherwise, the use of a regular quantizer results in a distortion penalty, as illustrated
			in Example~\ref{ex:scalar:energy}.
	\item One key contribution of this theorem is the additional tail condition for infinite-support
			source densities, which effectively limits the distortion contribution in the overload region. 
			This generalizes the class of probability densities for which distortion can be stably
			bounded using high-resolution approximations~\cite{BucklewW:82,CambanisG:83,Linder:91}. 
			We will demonstrate this with quantization of a Cauchy-distributed scalar 
			in Example~\ref{ex:scalar:cauchy}.
	\item For linear computations, the sensitivity is flat, meaning the optimal quantizer is the same as in the 	
			MSE-optimized case. 
			Hence, functional theory will lead to new quantizer designs only when the computation is nonlinear.
	\item In the proof of Theorem~\ref{thm:uvdist}, the first mean-value theorem is used on 
			both $f_X$ and $\lambda$, implying these densities are continuous. 
			However, this requirement can be loosened to piecewise-continuous distributions provided the tail
			conditions still hold and a minor adjustment is made on how partition boundaries are 
			chosen~\cite{CambanisG:83}. 
			Rather than elaborating further, we refer the reader to a similar extension 
			in~\cite[Sec. III-F]{MisraGV:11}.
			An equivalent argument can also be made for $g$ having a finite number
			of discontinuities in its first and second derivatives.
	\item The theorem assumes that $g$ is continuous and differentiable on the granular region, 
			meaning the sensitivity is positive.
			However, for explicit regions where $g^\prime(x) = 0$, the use of ``don't care'' regions 
			can be used to relax these conditions \cite[Sec. VII]{MisraGV:11}.
\end{asparaenum}

\subsection{Asymptotically Optimal Quantizer Sequences}
\label{sec:uv:optimal}

Since the fMSE of Theorem~\ref{thm:uvdist} matches~\eqref{eq:fmse:dist}, 
the optimizing quantizers are the same.
Using the recipe of Sec.~\ref{sec:prelim:hrsq}, 
we can show the optimal point density for fixed-rate quantization is
\begin{equation}
	\label{eq:dfsq2:lamfr}
	\lambda^*_{\mathrm{fmse,fr}}(x) = 
			\frac{\left( \gamma^2(x) f_X(x) \right)^{1/3}}
							{\int \left( \gamma^2(t) f_X(t) \right)^{1/3} \, dt} ,
\end{equation}
with distortion
\begin{equation}
	\label{eq:dfsq2:distfr}
	D^*_{\mathrm{fmse,fr}}(R) \APPROX \frac{1}{12} \, \| \gamma^2 f_X \|_{1/3} \, 2^{-2R} .
\end{equation}

Meanwhile, optimization in the entropy-constrained case yields
\begin{equation}
	\label{eq:dfsq2:lamvr}
	\lambda^*_{\mathrm{fmse,ec}}(x) = \frac{\gamma(x)}{\int \gamma(t) \, dt} ,
\end{equation}
giving distortion
\begin{equation}
	\label{eq:dfsq2:distvr}
	D^*_{\mathrm{fmse,ec}}(R) \APPROX \frac{1}{12} \, 2^{2 h(X) + 2\E[\log \gamma(X)]} \, 2^{-2R} .
\end{equation}

\subsection{Negligible Suboptimality of Simple Decoder}
\label{sec:uv:subopt}
\begin{figure}
	\centering
	\begin{tabular}{c}
	\includegraphics[width=3in]{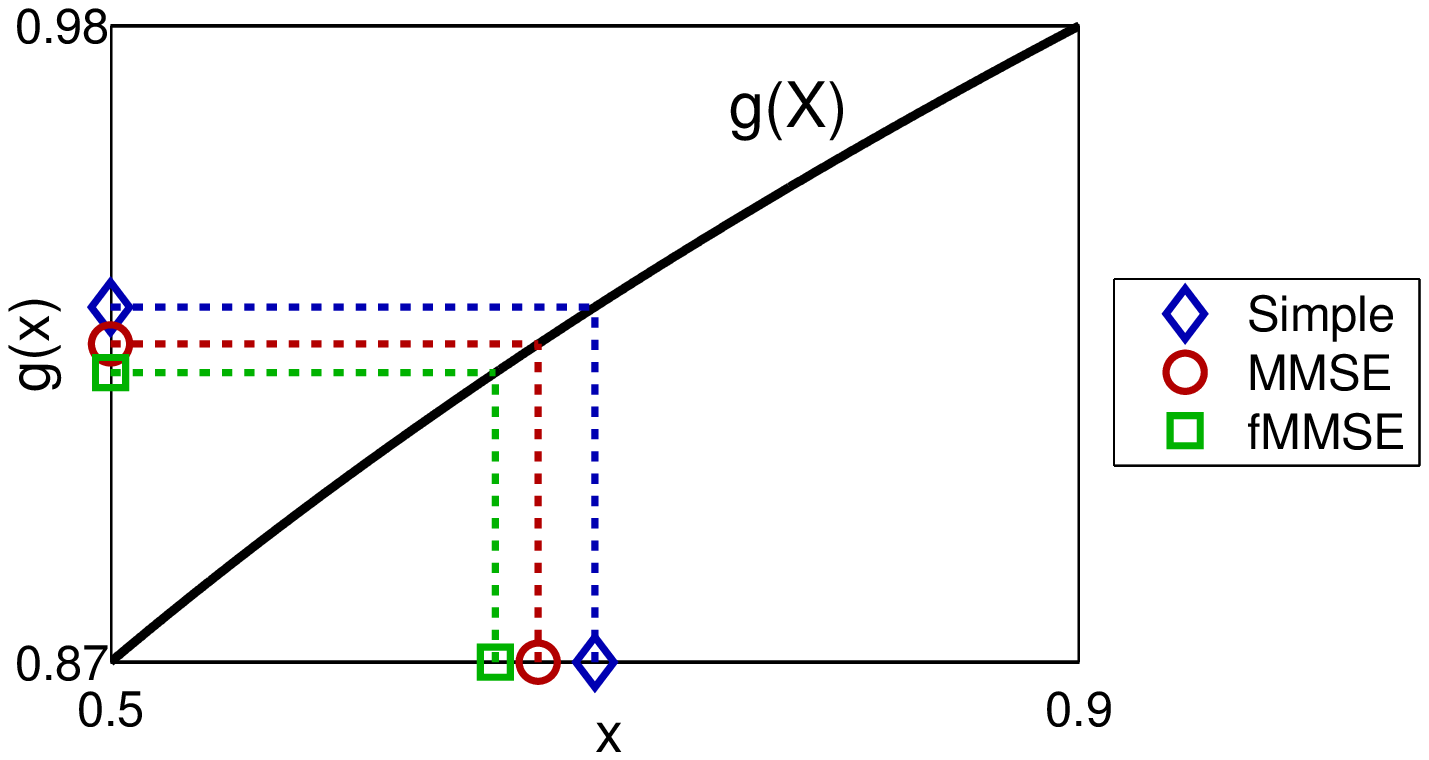} \\
	{\small (a)} \\
	\includegraphics[width=3in]{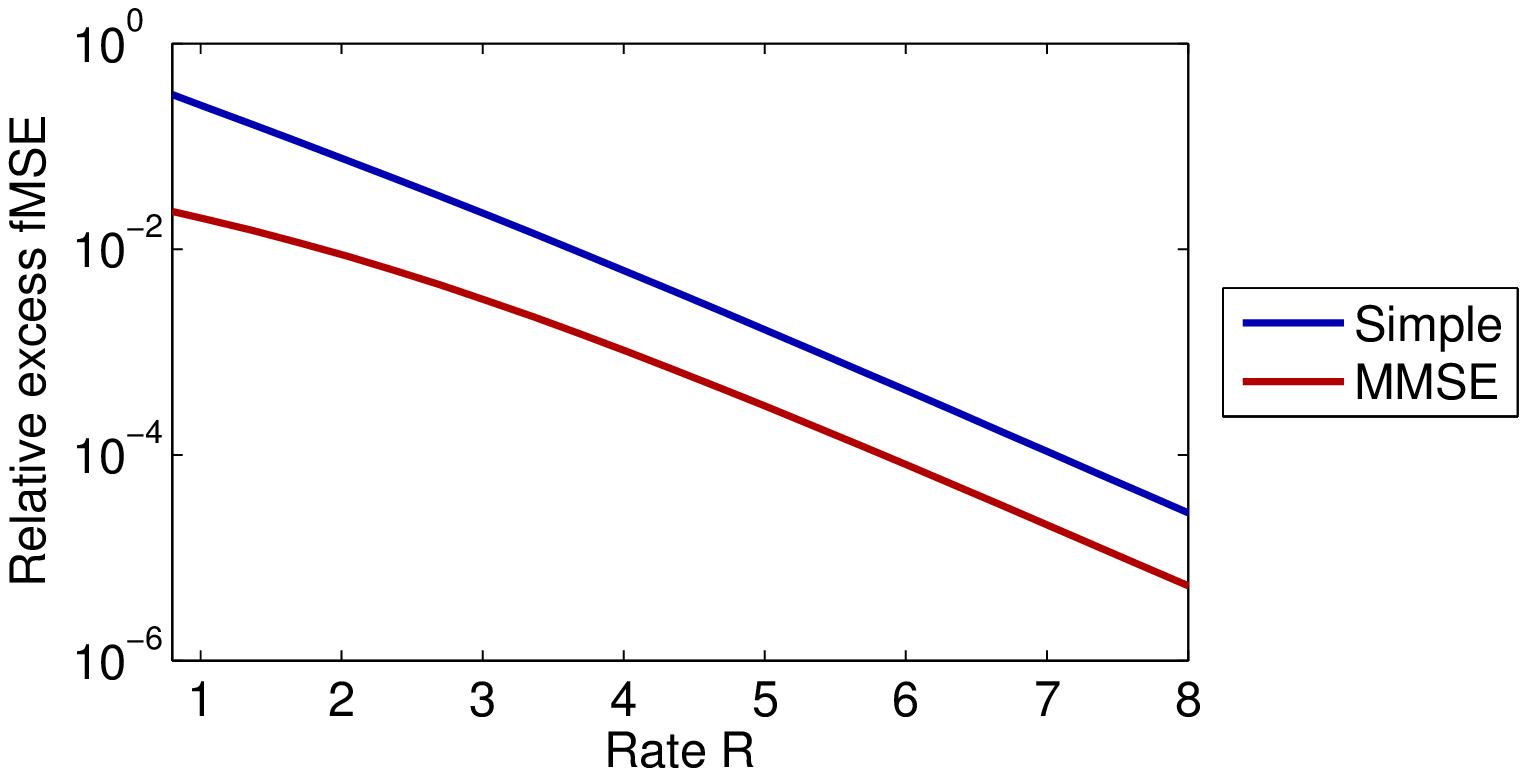} \\
	{\small (b)}
	\end{tabular}
	\caption{ 
	  	(a) Codeword placement under simple, MMSE, and fMMSE decoders.
	  				The simple decoder performs midpoint reconstruction followed by the application of
	  				the computation $g$. 
	  				The MMSE decoder applies $g$ to the conditional expectation of $X$ within the cell.
	  				Finally, the fMMSE decoder determines \eqref{eq:optdec} for the cell.
	  				In this example, the source distribution is exponential and the computation is concave. 
	  	(b) Performance loss due to the suboptimal codeword placement with respect to rate. 
	  				We can see that relative excess fMSE decreases linearly with rate and hence the fMSE of the 
	  				resulting quantizers are asymptotically equivalent. }
	\label{fig:codeword}
\end{figure}

Recall the simple decoder analyzed in this work is the computation $g$ applied to midpoint reconstruction as formulated in
\eqref{eq:midpointReconstruction}. 
One can do better by applying $g$ after finding the conditional MMSE estimate of $X$ utilizing knowledge of the 
source distribution only, 
or the fMMSE estimator \eqref{eq:optdec} incorporating the function as well. 
The codeword placements of the three decoders are visualized through an example in Fig.~\ref{fig:codeword}(a).
The asymptotic match of the performance of the simple decoder to the optimal
estimator \eqref{eq:optdec} is a main contribution of this paper.

The simple decoder is suboptimal because it does not consider the source distribution at all,
or equivalently assumes the distribution is uniform and the sensitivity is constant over the cell. 
High-resolution analysis typically approximates the source distribution as uniform over small cells~\cite{GrayN:98}, and the proof of Theorem~\ref{thm:uvdist}
utilizes the fact that the sensitivity is approximately flat over very small regions as well.
Hence, the performance gap between the simple decoder and the fMMSE estimator becomes negligible in
the high-resolution regime. 

To illuminate the rate of convergence, we study the performance gap as a function of 
quantization cell width, which is dependent on the communication rate (Fig.~\ref{fig:codeword}(b)).
We see the the relative excess fMSE (defined as 
$(D_\mathrm{dec} - D_\mathrm{opt})/D_\mathrm{opt}$) is exponential in rate, meaning 
\begin{equation}
	\frac{D_\mathrm{simple}}{D_\mathrm{opt}} = 1+ c_1 e^{-c_2 R}
\end{equation}
for some constants $c_1$ and $c_2$.
The speed at which the performance gap shrinks contributes greatly to why the high-resolution theory is
successful even at low communication rates. 

\section{Multivariate Functional Quantization}
\label{sec:mv}

We now describe the main result of the paper for the scenario shown in Fig.~\ref{fig:star},
where $N$ random scalars $(X_1, \ldots , X_N)$ are individually quantized and a scalar computation 
$g(\hat{X}_1^N)$ is performed.  
Assume the following conditions on the multivariate joint density, computation and quantizers
over a granular region $S_K \subset \mathbb{R^N}$:
\begin{asparaenum}[MF1$^\prime$.]
	\item The joint pdf $f_{X_1^N}$ is continuous and always positive on $S_K$ for any finite $K$.
	\item The multivariate function $g$ is continuous and twice differentiable in every 
				argument over $S_K$. 
				Every first- and second-order derivative is uniformly bounded by a constant $C_m$.
	\item For any $i,j \in \{1,\ldots,n\}$ and $m=0,1,2$, 
				$f_{X_i,X_j}(x_i,x_j) \gamma_i^m(x_i) \lambda_i^{-1}(x_i) \lambda_j^{-1}(x_j)$ is 
				Riemann integrable over $S_K$.
	\item We adopt the notation $x_{\backslash n}$ for $x_1^N$ with the $n$th element removed;
			an inverse operator
			$\tilde{x}(x_n, x_{\backslash n})$ outputs a length-$N$ vector with $x_n$ in the $n$th element.
			Then for every index $n$,
			the following holds for every $x_{\backslash n}$:
			\begin{align*}
				& \lim_{y \to \infty} \frac{ \int_y^\infty 
						|g(\tilde{x}(x, x_{\backslash n})) - g(\tilde{x}(y, x_{\backslash n}))|^2 
						f_{X_1^N}(\tilde{x} (x, x_{\backslash n})) \, dx}
						{\left( \int_y^\infty \lambda_n(x) \, dx \right)^2} \\
				& \hspace{4ex} = 0 .
			\end{align*}
			An analogous condition holds for the corresponding negative-valued tails.
\end{asparaenum}

Recalling $\QuantV$ and $\Lv$ represent a set of $N$ quantizers and point densities 
respectively, we present a theorem similar to Theorem~\ref{thm:uvdist}:
\begin{thm}
	\label{thm:mvdist}
	Assume $f_{X_1^N}$, $g$, and $\Lv$ satisfy conditions MF1$^\prime$--MF4$^\prime$.
	Also assume a fractional allocation $\alpha_1^N$  such that every $\alpha_n > 0$ and $\sum_n \alpha_n = 1$,
	meaning a set of quantizers $\QuantV$ will have $K_n = \alpha_n \kappa$ for some total allocation $\kappa$.	
	Then the fMSE
	\begin{equation}
		\label{eq:mvdist}
		\Dv = \E \left[ | g(X_1^N) - g(\QuantV(X_1^N)) |^2 \right]
	\end{equation}
	of this distributed system has the following limit:
	\begin{equation}
		\label{eq:mvdistHR}
		\lim_{\kappa \to \infty} \kappa^2 \Dv 
				= \sum_{n=1}^N \frac{1}{12 \alpha_n^2} \E \left[ \left(\frac{\gamma_n(X_n)}{\lambda_n(X_n)} 
							\right)^2 \right] .
	\end{equation}
\end{thm}

\begin{IEEEproof}
	See Appendix~\ref{app:thm:mvdist}.
\end{IEEEproof}

\subsection{Remarks}
\label{sec:mv:remarks}

\begin{asparaenum}[1.]
	\item Like in the univariate case, the simple decoder has performance that is asymptotically equivalent
			to the more complicated optimal decoder~\eqref{eq:optdec}.
	\item Here, the computation cannot generally be performed before quantization because encoders are distributed. 
			The exception is when the computation is \emph{separable}, meaning it can be decomposed
			into a linear combination of computations on individual scalars. 
			As a result, the sensitivity is no longer a conditional expectation and quantizer design simplifies to
			the univariate case, as demonstrated in Example~\ref{ex:multi:gauss}.
	\item The strict requirements of MF1$^\prime$ and MF2$^\prime$ could potentially be loosened. 
			However, simple modification of individual quantizers like in the univariate case is insufficient 
			since discontinuities may lie on a manifold that is not aligned with the partition lattice 
			of the $N$-dimensional space. 
			As a result, the error from using a planar approximation through Taylor's Theorem will be
			$\mathcal{O}(1/\kappa)$, which is no longer negligible. 
			However, based on experimental observations, such as in Example 4, we believe
			that when these discontinuities exist on a manifold of Jordan measure zero their error may be accounted for.
			Techniques similar to those in the proofs from~\cite{MisraGV:11} could potentially be useful in showing
			this rigorously.
\end{asparaenum}

\subsection{Asymptotically Optimal Quantizer Sequences}
\label{sec:mv:optimal}

As in the univariate case, the optimal quantizers match those in previous DFSQ work since the distortion
equations are the same. 
Using H\"{o}lder's inequality, the optimal point density for fixed-rate quantization for each source $n$ (communicated with 
rate $R_n$) is
\begin{equation}
	\label{eq:dfsq3:lamfr}
	\lambda^*_{n,\mathrm{fmse,fr}}(x) = 
			\frac{\left( \gamma_n^2(x) f_{X_n}(x) \right)^{1/3}}
							{\int_{-\infty}^\infty \left( \gamma_n^2(t) f_{X_n}(t) \right)^{1/3} \, dt} , 
\end{equation}
with fMSE
\begin{equation}
	\label{eq:dfsq3:distfr}
	D^*_{\mathrm{fmse,fr}}(R_1^N) \APPROX \frac{1}{12} \sum_{n=1}^N \| \gamma_n^2 f_{X_n} \|_{1/3} \, 2^{-2R_n} .
\end{equation}
Similarly, the best point density for the entropy-constrained case is
\begin{equation}
	\label{eq:dfsq3:lamvr}
	\lambda^*_{n,\mathrm{fmse,ec}}(x)  = \frac{\gamma_n(x)}{\int_{-\infty}^\infty \gamma_n(t) \, dt} , 
\end{equation}
leading to a fMSE of 
\begin{equation}
	\label{eq:dfsq3:distvr}
	D^*_{\mathrm{fmse,ec}}(R_1^N) \APPROX \frac{1}{12} \sum_{n=1}^N 2^{2 h(X_n) + 
								2\E[\log \gamma(X_n)]} \, 2^{-2R_n} .
\end{equation}

The rate allocations in~\eqref{eq:dfsq3:distfr} and~\eqref{eq:dfsq3:distvr} are allowed to vary. 
Given a total communication rate $R$, the optimal choice of $R_1^N$ is known~\cite{MisraGV:11,Segall:76}.

\subsection{Vector-valued Functions}
\label{sec:mv:vecfunc}

In Theorem~\ref{thm:mvdist}, we assumed the computation $g$ is scalar-valued.  
For completeness, we now consider vector-valued functions, where the output of $g$ is a vector in $\mathbb{R}^M$.
Here, the distortion measure is a weighted fMSE:
\begin{align}
	& D_\mathrm{fmse} (\Kv, \Lv, \beta_1^M) \\
	&		\hspace{2ex} = \sum_{m = 1}^M \beta_m  \E \left[ | g^{(m)}(X_1^N) - g^{(m)}(\QuantV(X_1^N)) |^2 \right] , \notag
\end{align}
where $\beta_1^M$ is a set of scalar weights and $g^{(m)}$ is the $m$th entry of the output of $g$. 
Through a natural extension of the proof of Theorem~\ref{thm:mvdist}, we can find the limit of the
weighted fMSE assuming each entry of the vector-valued function satisfies MF1$^\prime$--MF4$^\prime$.
	\begin{cor}
	\label{cor:mvdistVec}
	The weighted fMSE of a source $f_{X_1^N}$, computation $g$, set of scalar quantizers $\QuantV$, and 
	fractional allocation $\alpha_1^N$ has the following limit:
	\begin{equation}
		\lim_{\kappa \to \infty} \kappa^2 D_\mathrm{fmse} (\Kv, \Lv, \beta_1^M) 
				= \sum_{n=1}^N \frac{1}{12 \alpha_n^2} \E \left[ \left(\frac{\gamma_n(X_n,\beta_1^M)}{\lambda_n(X_n)} 
							\right)^2 \right] ,
	\end{equation}
	where the sensitivity profile is
	\begin{equation}
		\gamma_n(x,\beta_1^M) = \left( \sum_{m=1}^M \beta_m \E \left[ |g^{(m)}_n(X_1^N)|^2 \, | \, X_n = x \right]
						 \right)^{1/2} .
	\end{equation}
\end{cor}

\section{Examples}
\label{sec:examples}

In this section, we present examples for both univariate and multivariate functional quantization
using asymptotic expressions and empirical results from sequences of real quantizers. 
The empirical results are encouraging since the convergence to asymptotic limits is fast, 
usually when the quantizer rate is about 4 bits per source variable. 
This is because the Taylor remainder term in the distortion calculation decays with an extra
$\kappa$ factor, which is exponential in the rate.

\subsection{Examples for Univariate Functional Quantization}
\label{sec:examples:uv}

Below we present two examples of functional quantization in the univariate case. 
The theoretical results follow directly from Sec.~\ref{sec:uv}.

\begin{figure}
  \begin{center}
    \includegraphics[width=3in]{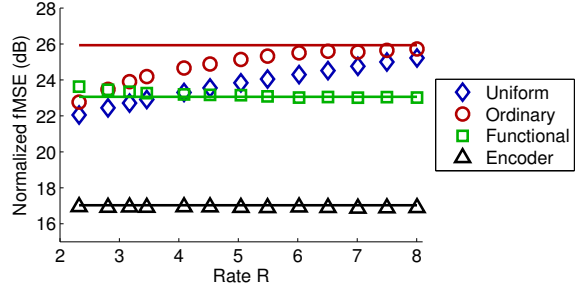}
  \end{center}
  \caption{Empirical and theoretical performance for the uniform, ordinary and functional quantizers, 
  		as well as the case when the computation is performed first.
			The distribution is standard normal and $g(x) = x^2$. 
			The distortions are multiplied by $2^{2 R}$ to better indicate the convergence results.}
  \label{fig:ex1}
\end{figure}

\begin{ex}
	\label{ex:scalar:energy}
	Assume $X \sim \mathcal{N}(0,1)$ and $g(x) = x^2$, yielding a sensitivity profile $\gamma(x) = 2 |x|$.
	We consider uniform quantizers, optimal ``ordinary'' quantizers
	(quantizers optimized for distortion of the source variable rather than the computation)
	given in Sec.~\ref{sec:prelim:hrsq},
	and optimal functional quantizers given in Sec.~\ref{sec:uv:optimal}, for a range of rates.
	The point densities of these quantizers, the source density $f_X$, and computation $g$ satisfy 
	UF1$^\prime$-UF4$^\prime$ and hence we utilize Theorem~\ref{thm:uvdist} to find asymptotic 
	distortion performance. 
	We also design practical quantizers for a range of $R$ and find the empirical 
	fMSE through Monte Carlo simulations using a random Gaussian source.
	In the fixed-rate case, theoretical and empirical performance are shown (Fig.~\ref{fig:ex1}).
	
	The distortion-minimizing uniform quantizer has a granular region that depends on $R$,
	which was explored in~\cite{HuiN:01}.
	Here, we simply perform a brute-force search to find the best granular region and the corresponding 
	distortion.
	Surprisingly, this choice of the uniform quantizer performs better over moderate rate regions than the 
	MSE-optimized quantizer. 
	This is because the computation is less meaningful where the source density is most likely and the MSE-optimized
	quantizer places most of its codewords. 
	Hence, one lesson from DFSQ is that using standard high-resolution theory may yield \emph{worse}
	performance than a naive approach for some computations. 
	Meanwhile, the functional quantizer optimizes for the computation and gives an additional 3 dB gain
	over the optimal ordinary quantizer.
	There is still a loss in using regular quantizers due to the computation being non-monotonic. 
	In fact, if the computation can be performed prior to quantization, we gain an extra bit for encoding 
	the magnitude and thus 6~dB of performance. 
	This illustrates Remark 2 of Sec.~\ref{sec:uv:remarks}.
	
	In the fixed-rate case, the empirical performance approaches the distortion limit described by
	Theorem~\ref{thm:uvdist}. 
	The convergence is fast and the asymptotic results predict practical quantizer performance at rates
	as low as 4 bits/sample. 
\end{ex}

\begin{figure}
  \begin{center}
    \includegraphics[width=3in]{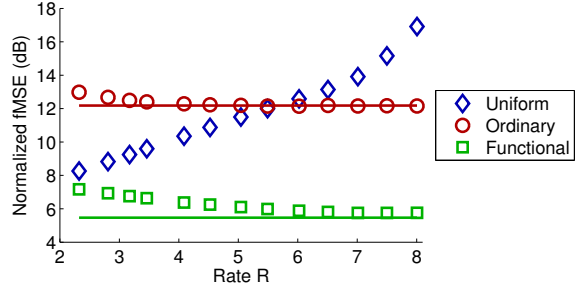}
  \end{center}
  \caption{Empirical and theoretical performance for the uniform, ordinary and functional 
  quantizers for Cauchy $f_X$ and $g(x) = e^{-|x|}$. 
  The distortions are multiplied by $2^{2 R}$ to better indicate the convergence results.}
  \label{fig:ex2}
\end{figure}

\begin{ex}
	\label{ex:scalar:cauchy}
	Let a source $X$ be distributed according to the Cauchy distribution centered around 0. 
	This heavy-tail density is special in that the mean and all higher moments are not defined. 
	Hence, it does not satisfy the conditions needed for high-resolution theory previously
	specified in~\cite{BucklewW:82,CambanisG:83,Linder:91}.
	However, the functional distortion can be asymptotically determined assuming Condition UF4$^\prime$ is satisfied. 
	The computation $g(x) = \operatorname{exp}(-|x|)$ and the Cauchy density satisfy UF4$^\prime$, and
	we confirm that experimental results match the theoretical computation of asymptotic distortion in 	
	Fig.~\ref{fig:ex2}.
\end{ex}

\subsection{Examples for Multivariate Functional Quantization}
\label{sec:examples:mv}

We next provide two examples that follow from the theory of Sec.~\ref{sec:mv}. 

\begin{ex}
	\label{ex:multi:gauss}
	Let $N$ sources be iid standard normal random variables and the computation be $g(x_1^N) = \sum_n x_n^2$. 
	Since the computation is separable, the sensitivity profile of each source is $\gamma_n(x) = |x|$,
	and the quantizers are the same as in Example~\ref{ex:scalar:energy}. 
	The distortion is also the same, except now scaled by $N$. 
\end{ex}

\begin{figure}
  \begin{center}
    \includegraphics[width=3in]{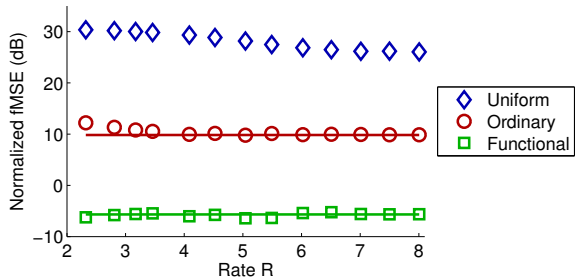}
  \end{center}
  \caption{Empirical and theoretical performance are provided for the uniform, ordinary and functional 
  				quantizers for $N=10$, exponential $f_X$, and $g(x_1^N) = \min(x_1^N)$.
  				The distortions are multiplied by $2^{2 R}$ to better indicate the convergence results.}
  \label{fig:ex4}
\end{figure}

\begin{ex}
	\label{ex:multi:min}
	Let $N$ sources be iid exponential with parameter $\lambda = 1$ and the computation be
	$g(x_1^N) = \min(x_1^N)$.
	In this case, Condition MF2$^\prime$ is not satisfied since there exists $N(N-1)/2$ two-dimensional
	planes where the derivative is not defined. 
	However, as discussed in the remarks of Theorem~\ref{thm:mvdist}, we strongly suspect
	we can disregard the distortion contributions from these surfaces.
	The overall performance, ignoring the violation of condition MF2$^\prime$, may be analyzed using the sensitivity:
	\begin{align*}
		\gamma_n(x) &= \left( \E [ |g_n(X_1^N)|^2 \, | \, X_n = x] \right)^{1/2}\\
		&= \left( \operatorname{Pr}\{ \min(X_1^N) = X_1 \, | \, X_1 = x\} \right)^{1/2}\\
		&= (e^{-\lambda x})^{(N-1)/2} ,
	\end{align*}
	where the third line follows from the cdf of exponential random variables. 

	In Fig.~\ref{fig:ex4}, we experimentally verify that the asymptotic predictions are precise.  
	This serves as evidence that MF2$^\prime$ may be loosened.
\end{ex}

\section{Conclusions}
\label{sec:summary}

In this paper, we have extended distributed functional scalar quantization to a general class
of finite- and infinite-support distributions, and demonstrated that a simple decoder, performing the 
computation directly on the quantized measurements, achieves asymptotically equivalent performance to the fMMSE decoder.
Although there are some technical restrictions on the source distributions and computations to ensure
the high-resolution approximations are legitimate, 
the main goal of the paper is to show that 
DFSQ theory is widely applicable to distributed acquisition systems 
without requiring a complicated decoder. 
Furthermore, the asymptotic results give good approximations for the
performance at moderate quantization rates.

DFSQ has immediate implications in how sensors in acquisition networks collect and compress data 
when the designer knows the computation to follow. 
Using both theory and examples, we demonstrate that knowledge of the computation may change the 
quantization mapping and improve fMSE\@. 
Because the setup is very general, there is potential for impact in areas of signal acquisition where quantization is traditionally considered as a black box. 
Examples include multi-modal imaging technologies such as 3D imaging and parallel MRI\@. 
This theory can also be useful in collecting information for applications in machine learning and data mining. 
In these fields, large amounts of data are collected but the measure of interest is usually some
nonlinear, low-dimensional quantity. 
DFSQ provides insight on how data should be collected to provide more accurate results
when the resources for acquiring and storing information are limited.

\appendices

\section{Proof of Theorem~\ref{thm:uvdist}}
\label{app:thm:uvdist}

Taylor's theorem states that a function $g$ that is $n+1$ times 
continuously differentiable on a closed interval $[a,x]$ takes the form
\[ g(x) = g(a) + \left( \sum_{i=1}^n \frac{g^{(i)}(a)}{i!} (x-a)^i \right) + R_n(x,a) , \]
with a Taylor remainder term
\[ R_n(x,a) = \frac{g^{(n+1)}(\xi)}{(n+1)!} (x-a)^{n+1} \]
for some $\xi \in [a,x]$. 
More specific to our framework, for any $x \in [c_k, p_{k})$, the first-order remainder is bounded as
\begin{equation}
	\label{eq:taylorRbound_scalar}
	| R_1(x,c_k) | \leq \frac{1}{2} \max_{\xi \in [c_k, p_k]} 
							|g^{\prime\prime}(\xi)| (p_k - c_k)^2 . 
\end{equation}
Using Condition UF2$^\prime$, we will uniformly bound $|g^{\prime\prime}(\xi)|$ by $C_u$.

The first mean-value theorem for integrals states that for a continuous function
$r : [a,b] \to \mathbb{R}$ and integrable function $s : [a,b] \to [0,\infty)$ that does not change sign, 
there exists a value $x \in [a,b]$ such that
\beq \int_a^b r(t) s(t) \, dt = r(x) \int_a^b s(t) \, dt \mbox{.} \label{eq:MVT}\eeq
For the case of the companding quantizers, combining this with \eqref{eq:densityKRelation} means
\begin{equation}
	\label{eq:meanvaluequant}
	\frac{1}{K} = \int_{p_k}^{p_{k+1}} \lambda(x) \, dx = \lambda(y_k) (p_{k+1} - p_k) = \lambda(y_k) \Delta_k
\end{equation}
for some $y_k \in (p_k,p_{k+1}]$, where we have defined the $k$th quantizer cell length $\Delta_k = p_{k+1}-p_k$.
The relationship between $K$, $\lambda$, and $\Delta_k$ is central in the proof.

With these preparations, we continue to the proof. 
Consider expansion of $\D$ by total expectation:
\begin{equation}
	\label{eq:DKlambda_expansion}
 	\D = \sum_{k=0}^{K-1} \int_{p_k}^{p_{k+1}} 
 			| g(x) - g(c_k) |^2 
 			f_X(x) \, dx .
\end{equation}
We would like to eliminate the first and last terms of the sum because
the unbounded interval of integration would cause problems with the
approximation technique employed later.
The last term is
\begin{equation}
	\label{eq:DKlambda_lastterm}
	\int_{p_{K-1}}^{\infty} | g(x) - g(p_{K-1}) |^2 f_X(x) \, dx,
\end{equation}
where we have used $c_K = p_{K-1}$.
By Condition UF4$^\prime$, this is asymptotically negligible in comparison
to
\[ \left( \int_{p_{K-1}}^\infty \lambda(x) \, dx \right)^2 = \frac{1}{K^2}. \]
Thus the last term \eqref{eq:DKlambda_lastterm} does not contribute to
$\lim_{K\to\infty} K^2 \D$.
We can similarly eliminate the first term, yielding
\begin{align}
	K^2 \D
 		&\APPROX K^2 \sum_{k=1}^{K-2} \int_{p_k}^{p_{k+1}} 
 			| g(x) - g(c_k) |^2 
 			f_X(x) \, dx
 			\label{eq:thm1:distsummands}.
\end{align}
Effectively, Condition UF4$^\prime$ promises that the tail of the source distribution is decaying fast enough
that we can ignore the distortion contributions outside the extremal codewords.

Further expansion of~\eqref{eq:thm1:distsummands} using Taylor's theorem with remainder yields: 
\begin{align*}
	\allowdisplaybreaks
	&\lefteqn{K^2 \D} \\
	& \hspace{1ex} \APPROX
		K^2 \sum_{k=1}^{K-2} 
		\int_{p_k}^{p_{k+1}} 
		| g'(c_k)(x-c_k) + R_1(x,c_k) |^2 
		f_X(x) \, dx 
		\\
	& \hspace{1ex} = \underbrace{K^2 \sum_{k=1}^{K-2} \int_{p_k}^{p_{k+1}} 
		|g'(c_k)|^2 \, |x-c_k|^2 
		f_X(x) \, dx}_{\text (A)} 
		\\
	& \hspace{2ex} + \underbrace{K^2 \sum_{k=1}^{K-2} 2 \int_{p_k}^{p_{k+1}}
		|R_1(x,c_k)| \, |g'(c_k)| \, |x-c_K| 
		f_X(x) \, dx}_{\text (B)}
		\\
	& \hspace{2ex} + \underbrace{K^2\sum_{k=1}^{K-2} \int_{p_k}^{p_{k+1}}
		R_1(x,c_k)^2 
		f_X(x) \, dx}_{\text (C)}  
		\mbox{.}
\end{align*}
Of the three terms, only term (A) has a meaningful contribution. 
It can be simplified as follows:
\begin{align}
	\allowdisplaybreaks
	& K^2 \sum_{k=1}^{K-2} \int_{p_k}^{p_{k+1}} 
		| g^\prime(c_k)|^2 \, |x - c_k|^2 
		f_X(x) \, dx 
		\notag \\
	& \hspace{2ex} \stackrel{(a)}{=} K^2 \sum_{k=1}^{K-2} 
		| g^\prime(c_k) |^2 f_X(v_k) 
		\int_{p_k}^{p_{k+1}} |x-c_k|^2 \, dx 
		\notag \\
	& \hspace{2ex} \stackrel{(b)}{=} K^2 \sum_{k=1}^{K-2}
		| g^\prime(c_k) |^2 f_X(v_k) \frac{\Delta_k^3}{12} 
		\notag \\
	& \hspace{2ex} \stackrel{(c)}{=} \frac{1}{12} \sum_{k=1}^{K-2} f_X(v_k) 
		\left( \frac{|g^\prime(c_k)|^2}{\lambda^2(y_k)} \right) \Delta_k  
		\notag \\
	& \hspace{2ex} \stackrel{(d)}{\longrightarrow} \frac{1}{12} \int_{S} 
		\left(\frac{g^\prime(x)}{\lambda(x)} \right)^2 
		f_X(x) \, dx 
		\label{eq:thm1:granular},
\end{align}
where 
(a) arises from using~\eqref{eq:MVT}, where $v_k$ is some point in the $k$th quantizer cell;
(b) is evaluation of the integral, recalling \eqref{eq:midpointReconstruction};
(c) follows from~\eqref{eq:meanvaluequant};
and (d) holds as $K \to \infty$ by the convergence of Riemann rectangles to the integral 
(assumption UF3$^\prime$).
 				
The higher-order error terms are negligible using the bound reviewed in~\eqref{eq:taylorRbound_scalar}.  
We now show that term (B) goes to zero:
\begin{align}
	\allowdisplaybreaks
	& K^2 \sum_{k=1}^{K-2} 2 \int_{p_k}^{p_{k+1}}
		|R_1(x,c_k)| \, |g'(c_k)| \, |x-c_K| 
		f_X(x) \, dx 
		\notag \\
	& \hspace{2ex} \stackrel{(a)}{\leq} K^2 \sum_{k=1}^{K-2} 
		C_u \Delta_k^2 |g^\prime(c_k)| 
		\int_{p_k}^{p_{k+1}}  |x-c_k| 
		f_X(x) \, dx 
		\notag \\
	& \hspace{2ex} \stackrel{(b)}{\leq} K^2 \sum_{k=1}^{K-2} 
		C_u \Delta_k^4 |g^\prime(c_k)| f_X(v_k) 
		\notag \\	
	& \hspace{2ex} \stackrel{(c)}{=} \frac{C_u}{K} \sum_{k=1}^{K-2} f_X(v_k) 
		\frac{|g^\prime(c_k)|}{\lambda^3(y_k)} \Delta_k 
		\notag \\
	& \hspace{2ex} \stackrel{(d)}{\longrightarrow} 0, 
		\label{eq:thm1:remainder}
\end{align}
where 
(a) follows from bounding $R_1(x, c_k)$ using~\eqref{eq:taylorRbound_scalar};
(b) arises from using~\eqref{eq:MVT} and bounding the integral;
(c) follows from \eqref{eq:meanvaluequant};
and (d) holds as $K \to \infty$ by the convergence of Riemann rectangles to the integral 
(assumption UF3$^\prime$). 
Hence, the distortion contribution becomes negligible as $K$ increases. 

A similar analysis can be used to show that expansion term (C) scales as $1/K^2$ with growing
codebook size and is therefore also negligible.

\section{Proof of Theorem~\ref{thm:mvdist}}
\label{app:thm:mvdist}

We parallel the proof of Theorem~\ref{thm:uvdist} using Taylor expansion 
and bounding the distortion contributions of each cell. 
We review the first-order version of the multivariate Taylor's theorem: 
a function that is twice continuously differentiable on a closed ball $B$ takes the form
\[ g(x_1^N) = g(a_1^N) +  \sum_{n=1}^N 
				\left[ g_n(a_1^N) (x_n - a_n)\right] + 
				R_1(x_1^N, a_1^N) , \]
where we recall that $g_n(x_1^N)$ is the partial derivative of $g$ with respect to the $n$th argument evaluated at the point $x_1^N$.
The remainder term is bounded by
\[ R_1(x_1^N, a_1^N) \leq \sum_{i=1}^N \sum_{j=1}^N (x_i-a_i) (x_j-a_j) C_m ,\]
under Condition MF2$^\prime$.   
Applying a linear approximation to a quantizer cell $s$ with midpoint $(c_s)_1^N$ 
and side lengths $\{\Delta_i(s)\}_{i=1}^N$, the Taylor residual is
\begin{align}
	| R_1\left(x_1^N, (c_s)_1^N\right) | & \leq C_m \sum_{i=1}^N \sum_{j=1}^N \Delta_i \Delta_j \notag \\
			& \stackrel{\text{(a)}}{\leq} N C_m \overline{\Delta} \sum_{i=1}^N \Delta_i \notag \\
			& \stackrel{\text{(b)}}{=} NC_m \overline{\Delta} \sum_{i=1}^N 
									\frac{1}{\lambda_i(y_{s,i}) \kappa \alpha_i} \notag \\
			& \stackrel{\text{(c)}}{\leq} \frac{NC_m \overline{\Delta}}{\kappa \underline{\alpha}} 
									\sum_{i=1}^N \lambda_i(y_{s,i})^{-1} \mbox{,} \label{eq:multiRemainderBound}
\end{align}
where in (a) we define $\overline{\Delta}$ as the longest quantizer interval length in any dimension;
in (b) we invoke \eqref{eq:meanvaluequant} with $y_{s,i}$ being the $i$th coordinate of some point 
in quantizer cell $s$;
and in (c) we define $\underline{\alpha}$ as the smallest $\alpha_i$.

Let $\mathcal{S}_K$ be the partition lattice induced by $N$ scalar quantizers, excluding the overload regions.
By total expectation, we find the distortion of each partition cell and sum their contributions.
By Condition MF4$^\prime$, the distortion from overload cells become negligible with increasing $\kappa$
and can be ignored. 
Using Taylor's theorem, the scaled total distortion becomes
\[
\kappa^2 \Dv = A + B + C,
\]
where
\begin{align*}
 A & = \kappa^2 \sum_{s \in \mathcal{S}_K} \int_{x_1^N \in s} \sum_{i=1}^N \sum_{j=1}^N
				g_i\left((c_s)_1^N\right) g_j\left((c_s)_1^N\right) \\
   & \qquad \qquad \qquad \quad \cdot (x_i - c_{s,i}) (x_j - c_{S,j}) 
				f_{X_1^N}(x_1^N) \, d x_1^N , \\
 B & = \kappa^2 \sum_{s \in \mathcal{S}_K} 
				\int_{x_1^N \in s} 
				2\sum_{n=1}^N |g_n\left((c_s)_1^N\right)| \, |x_n - c_{s,n}| \\
   & \qquad \qquad \qquad \quad \cdot |R_1\left(x_1^N, (c_s)_1^N\right)|
				\, f_{X_1^N}(x_1^N) \, d x_1^N , \\
 C & = \kappa^2 \sum_{s \in \mathcal{S}_K} 
				\int_{x_1^N \in s} 
				R_1^2\left(x_1^N, (c_s)_1^N\right) 
				f_{X_1^N}(x_1^N) \, d x_1^N.
\end{align*}

In term $A$, we may disregard all cross terms since $(X_n - c_{k,n})$ becomes uncorrelated in 
the high-resolution approximation because the pdf in each cell becomes well-approximated by a 
uniform distribution as the cell gets smaller.
The remaining components of the distortion are
\[ \kappa^2 \sum_{s \in \mathcal{S}_K} 
			\int_{x_1^N \in s} 
			\left( \sum_{n=1}^N g^2_n\left((c_s)_1^N\right) (x_n - c_{s,n})^2 \right) 
			f_{X_1^N}(x_1^N) \, d x_1^N . \]
Using~\eqref{eq:meanvaluequant}, the distortion contribution becomes
\begin{equation*}
	\kappa^2 \sum_{s \in \mathcal{S}_K}
			\int_{x_1^N \in s} 
			\left( \sum_{n=1}^N  \frac{g^2_n\left((c_s)_1^N\right)}{12 K_n^2 \lambda^2_n(y_{s,n})} \right) 
			f_{X_1^N}(x_1^N) \, d x_1^N \mbox{,}
\end{equation*}
where $y_{s,n}$ is the $n$th coordinate of some point in quantizer cell $s$.
Using assumption MF3$^\prime$, this approaches the integral expression
\begin{eqnarray*}
	\lefteqn{\sum_{n=1}^N \frac{1}{12 \alpha_n^2}
						\E \left[ \left(\frac{g_n\left((c_s)_1^N\right)}{\lambda_n(X_n)}\right)^2 \right] } \\
	 & = & \sum_{n=1}^N \frac{1}{12 \alpha_n^2}
						\E \left[ \left(\frac{\gamma_n(X_n)}{\lambda_n(X_n)}\right)^2 \right] \mbox{.}
\end{eqnarray*}

It remains to show that the remainder terms $B$ and $C$ may be ignored.  
First, consider any of the $N$ summands that constitute $B$:
\begin{align*}
	& \kappa^2 \sum_{s \in \mathcal{S}_K} 
		\int_{x_1^N \in s} 
		2 | g_n\left((c_s)_1^N\right) | \, |x_n - c_{s,n}| \\
        & \qquad \qquad \qquad \cdot |R_1\left(x_1^N,(c_s)_1^N\right)| \,
		f_{X_1^N}(x_1^N) \, d x_1^N
		\\
	& \hspace{2ex} \stackrel{(a)}{\leq} \kappa^2 \sum_{s \in \mathcal{S}_K} 
		\int_{x_1^N \in s} 
		2 | g_n\left((c_s)_1^N\right) | \, |x_n - c_{s,n}| \\
        & \qquad \qquad \qquad \cdot
		\frac{{N} C_m \overline{\Delta}}{\underline{\alpha} \kappa} \sum_{i=1}^N \lambda_i(y_{s,i})^{-1} 
		f_{X_1^N}(x_1^N) \, d x_1^N
    \\ 
	& \hspace{2ex} \stackrel{(b)}{\leq}  
		\frac{2 N C_m \overline{\Delta} \kappa}{\underline{\alpha}} \sum_{s \in \mathcal{S}_K} 
    |g_n\left((c_s)_1^N\right)| \Delta_n(s) \\
	& \qquad \qquad \qquad \cdot \sum_{i=1}^N \lambda_i(y_{s,i})^{-1}
    f_{X_1^N}\left((v_s)_1^N\right) \prod_{j=1}^N \Delta_j(s)
    \\  
	& \hspace{2ex} \stackrel{(c)}{\leq}
		\frac{2 N C_m \overline{\Delta}}{\underline{\alpha}} 
		\sum_{i=1}^N \sum_{s \in \mathcal{S}_K} 
    |g_n\left((c_s)_1^N\right)| \\
	& \qquad \qquad \cdot \lambda_i(y_{s,i})^{-1} \lambda_n(y_{s,n})^{-1}
	 	f_{X_1^N}\left((v_s)_1^N\right) \prod_{j=1}^N \Delta_j(s)
    \\  
	& \hspace{2ex} \stackrel{(d)}{\longrightarrow}  
		\left( \lim_{\kappa \rightarrow\infty} \frac{2 N C_m\overline{\Delta}} {\underline{\alpha}}\right) 
		\sum_{i=1}^N \E \left[ \frac{|g_n(X_1^N)|}{\lambda_i(X_i) \lambda_n(X_n)} \right]
    \\ 
	& \hspace{2ex} = 0 
		\mbox{,}
\end{align*}
where
(a) follows from \eqref{eq:multiRemainderBound};
(b) employs \eqref{eq:MVT} and bounds $|x_n - c_{s,n}| \leq \Delta_n$;
(c) invokes \eqref{eq:meanvaluequant}; 
and (d) is valid according to assumption MF3$^\prime$.

Remainder term $C$ is negligible in a similar manner, which proves the theorem.

\bibliographystyle{ieeetr}
\bibliography{newcond} 	

\end{document}